\documentstyle[11pt,newpasp,twoside,epsf]{article}
\def\edcomment#1{\iffalse\marginpar{\raggedright\sl#1\/}\else\relax\fi}
\reversemarginpar

\begin{document}
\title{Building Planets with Dusty Gas}
  \author{S. T. Maddison}
\affil{Centre for Astrophysics \& Supercomputing, Swinburne University,
       PO Box 218, Hawthorn, VIC 3122, Australia}
\author{R. J. Humble} 
\affil{CITA, University of Toronto, Canada and 
      Centre for Astrophysics \& Supercomputing, Swinburne University,
       Australia }
\author{J. R. Murray}
\affil{University of Leicester, University Road, Leicester LE1 7RH, UK}

\begin{abstract}
We have developed a new numerical technique for simulating dusty-gas
flows.  Our unique code incorporates gas hydrodynamics, self-gravity and
dust drag to follow the dynamical evolution of a dusty-gas medium.  We
have incorporated several descriptions for the drag between gas and dust
phases and can model flows with submillimetre, centimetre and metre size 
``dust".  We present calculations run on the APAC\footnote{Australian 
Partnership for Advanced Computing {\texttt{http://nf.apac.edu.au/}}} 
supercomputer following the evolution of the dust distribution in the 
pre-solar nebula.
\end{abstract}

\section{Introduction}

Up until recently we had only one observation to test our theories of planet 
formation against - our own Solar System. Now however with new planets and 
solar systems being both identified and parameterised at a rate approaching 
one a month, the observational constraints are much tighter and our lack of 
understanding of many aspects of the planet formation process is all too 
obvious.  At the most basic level we know that micron size grains of dust in 
the pre-solar nebula clump and coagulate together to form planets, objects 
10$^{13}$--10$^{14}$ times larger. Planet formation is a multi-stage process, 
taking us from dust grain to boulder to planetesimal to planetary embryo. 
Analytical arguments (Goldreich \& Ward 1973) have presented us with 
constraints on the time scales for each stage but little more. 
It is the very first stage of the process that we are concerned with in 
this paper -- from micron scale dust to metre sized boulders.  

Theoretical models have changed much in recent years and the simple picture 
of a thin dust layer accumulating at the disk midplane, becoming 
gravitationally unstable, and breaking into planetesimals (Safronov 1969; 
Goldreich \& Ward 1973) now seems unlikely.
Recently Goodman \& Pindor (2000) showed that turbulent drag causes radial 
instabilities in the dust layer, even if the disk self-gravity is negligible. 
In their steady state models, grains in a uniform dust layer experience 
radial drift as expected.  However, for perturbed disks they predict 
that over-dense rings form within an orbital period, with the ring thickness 
similar to the thickness of the dust layer.  These rings eventually collapse 
into planetesimals in the kilometre size range. 

In this paper we present the first three-dimensional numerical simulations 
that include the effects of hydrodynamical forces, self-gravity and gas drag 
upon an evolving dusty gas disk.  We describe a new numerical code, based 
upon the smoothed particle hydrodynamics (SPH) technique which uses a 
collection of particles to approximate a fluid. 
At present we run simulations with uniform grain size and do not allow 
dust particles to have individual or time varying chemical and physical 
properties. Adding this level of sophistication to the model would 
require only  minor changes to the code.

\section{The Planet Building Code}

We have developed a new parallel two-phase (dust \& gas) hydro $+$ tree 
code by merging a  dusty gas SPH code (Maddison 1998) with an MPI parallel 
Hashed Oct Tree N-body $+$ SPH code (Humble 1999).  Following  Monaghan 
\& Kocharyan (1995), the dusty gas is approximated by two inter-penetrating 
flows that interact via a drag force.

The particles are tagged as being gas or dust and are then allowed to interact 
according to gas-gas, gas-dust and dust-dust interactions.  Only gas 
particles feel a pressure force and are affected by viscosity.  The gas-dust 
combination feel the drag force and also the mixed term seen in both the dust 
and gas particle acceleration equations.  The only dust-dust interaction is 
through gravity.

The two fluids are coupled by gravity and drag, and the equation of motion is 
given by:
\begin{eqnarray*}
\frac{d {\bf v}_g}{dt} &=&  - \frac{1}{\rho}_g \nabla P + 
           \frac{K}{\hat{\rho}_g}({\bf v}_d-{\bf v}_g)  - \nabla \Phi \, ,\\ 
\frac{d {\bf v}_d}{dt} &=&  - \frac{1}{\rho}_d \nabla P - 
           \frac{K}{\hat{\rho}_d}({\bf v}_d-{\bf v}_g)  - \nabla \Phi \, ,
\end{eqnarray*}
where the terms on the right hand side are gas pressure, drag force and 
gravity respectively.  The density ${\hat{\rho}}$ is the mass density per 
unit volume, while $\rho$ is the local density (related via 
$ \hat{\rho} = \Theta \rho$, where $\Theta$ is the volumetric void fraction).  
The functional form of $K$ depends upon the drag regime (e.g. Epstein or 
Stokes) being considered.  In the simulations we present, we have 
implemented Epstein drag which is appropriate for protoplanetary disks. 
Thus $K = ({\rho \Theta C_{\rm drag} c})/{r_{\rm dust}}$, where $c$ the 
local sound speed, $C_{\rm drag}$ the drag coefficient, and $r_{\rm dust}$ 
the dust grain size.

The drag term is calculated using a pairwise implicit backward Euler scheme, 
while the hydrodynamics and self-gravity are solved explicitly.  Operator 
splitting is used to integrate particle orbits under the influence of the 
un-softened gravity of the central star.
The code has been tested for stability and accuracy with standard periodic 
box simulations to ensure that the drag terms are correct, and collapsing 
sphere simulations to test the gravity.  The code scales linearly to more 
than 32 processors.
For simplicity we assume that the dust grains do not evaporate or coagulate 
and that the gas does not condense. We take the dust grains to be 
incompressible.

\begin{figure}
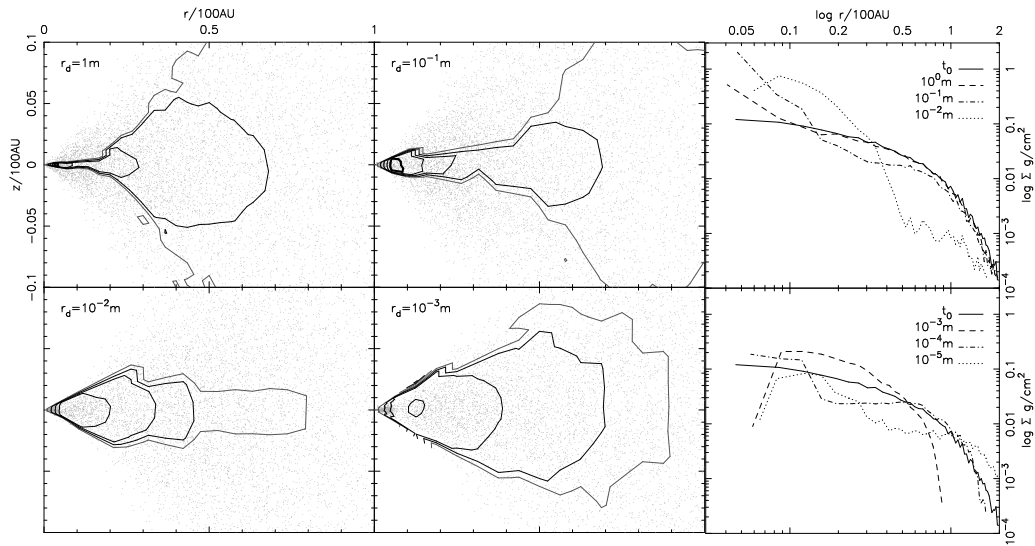

\plotfiddle{maddison-fig1.ps}{3cm}{0}{37}{37}{-200}{-110}
\plotfiddle{maddison-fig2.ps}{3cm}{0}{36.6}{36.6}{51}{-14}
\caption{
The 4 left hand panels show final gas (dots) and dust (contours) density 
distribution for 12.5K gas $+$ 12.5K dust models of various grain sizes.  Five 
contour levels (10$^{-13}$ to $10^{-17}$ g/cm$^3$) are plotted.  The 2 right 
hand panels show the final dust surface density distributions of all 6 models 
(initial distribution labelled t$_0$).}
\end{figure}

\section{The Simulations}

We are particularly interested in the effects of grain size upon disk
morphology and global disk dynamics. We present the results of 6 
simulations of gas disks laden with dust of different grain sizes.
All simulations start with a prolate spheroid of rotating, self-gravitating 
gas (with a star at the centre) that spins down to near-equilibrium.  The 
dust is then added (overlaid on the 3D flared gas disk) and the simulations 
are then followed for approximately 10$^4$ years (which corresponds to 
11 orbits at 100\,AU).

The model parameters used were: $M_{\star} = 1.0 \, M_{\sun}$,
$M_{\rm disk} = 0.01 \, M_{\star}$, $M_{\rm dust} = 0.01 \, M_{\rm disk}$ and
$R_{\rm disk} =$ 100\,AU.  The grain parameters were: 
$\rho_{\rm d} =$ 2.4\,g/cm$^3$ and  $r_{\rm d} = 1, 10^{-1}, 10^{-2}, 
10^{-3}, 10^{-4}$ and $10^{-5}$\,m.  The $\alpha$-disk parameters used
were: $c(R) = c_{\rm 0} (R/100AU)^{-3/8}$, $T(R) \propto (R/100AU)^{-3/4}$,
$H/R = 0.1$ at 100\,AU, and $\gamma = 5/3$ with an isothermal equation of 
state.  Low resolution runs used $2 \times 12500$ particles, and high  
resolution runs used $2 \times 125000$ particles.  The self-gravity softening 
$\varepsilon =$ 1.0\,AU, and the  SPH artificial viscosity parameters
$\alpha_{\rm SPH} = 0.1$ and $\beta_{\rm SPH} = 0.0$.

\section{Discussion}

For large (10m) and small (micron) dust sizes, we expect that the dust 
distribution will stay close to the initial flared disk. The largest grains 
(not shown) are weakly coupled to the gas, and if started in Keplerian motion, 
they will remain there.  On the other hand, the tiny grains are so strongly 
coupled to the gas that they are essentially co-moving (on the timescales we 
examine).  For both extremes, we see little evolution of the dust 
distribution. It is for the regime 0.1\,mm $< r_{dust} <$ 1\,m that the most 
significant disk evolution occurs.

In the $r-z$ plots of figure~1, significant deviation from the 
initially flared disk occurs in the inner regions of the 1m and 10cm plots, 
in the mid regions (from $r=0.6$ to $r=0.9$) of the 1cm plot, and in the 
outer regions of the 1mm and smaller plots.  
In these regions, the dust is moving at close to Keplerian speeds, whereas 
the gas is (as always) sub-Keplerian. As the velocity difference is 
sustained, the Epstein drag is optimal, and the energy loss rate is high 
-- allowing a thin layer of dust to form in the midplane and  migrate radially.
These thin dense dust 
disks are those that Goodman \& Pindor suggest have global turbulent 
instability modes. 

While the 10cm and 1cm dust exhibits the highest surface density (top right of 
figure~1), the volumetric density is largest in the inner regions of the 1m and 
10cm disks.  Therefore these size ranges are probably the most interesting 
from a planet formation viewpoint. 
        
The Lagrangian nature of the code means that it is trivial to add empirical 
grain growth models, and to follow the grain temperature and density 
histories and hence generate chemical compositions.  The equations of state 
and drag term can easily be altered on a per-region or per-particle basis to 
account for local disk conditions.

\acknowledgements
The research was supported by a Victorian Partnership for Advanced Computing 
Expertise Grant and an Australian Partnership for Advanced Computing Merit 
Allocation Grant.

\end{document}